\documentclass{epl}
\usepackage{amsmath}
\renewcommand{\epsilon}{\varepsilon}
\renewcommand{\phi}{\varphi}

\title{Universal Fluctuations of the Danube Water Level: a Link with Turbulence,
Criticality and Company Growth}
\author{Steven.T. Bramwell\inst{1}
\thanks{E-mail: \email{s.t.bramwell@ucl.ac.uk}} \and
Tom Fennell\inst{2} \and
Peter C. W. Holdsworth\inst{3}
\thanks{E-mail: \email{peter.holdsworth@ens-lyon.fr}}
\and Baptiste Portelli\inst{3}}

\institute{
\inst{1} University College London, Department of Chemistry, 20
Gordon Street, London WC1H~0AJ, UK.\\
\inst{2} The Royal Institution of Great Britain, 21
Albemarle Street, London W1X~4BS, UK.\\
\inst{3} Laboratoire de Physique, ENS-Lyon and CNRS, 
46 All\'ee d'Italie,
69007 Lyon, France.
}
\pacs{2.50 r}{Probability theory, stochstic processes, statistics}
\pacs{5.40 -a}{Fluctuation phenomena, random processes, noise, Brownian motion}
\pacs{5.65+b}{Self-organized systems}

\begin{document}

\maketitle

\begin{abstract}
A global quantity, regardless of its precise nature, will often
fluctuate according to a Gaussian limit distribution. However, in highly
correlated systems, other limit distributions are possible. We have
previously calculated one such distribution and have argued that this
function should apply specifically, and in many instances, to global
quantities that define a steady state~\cite{BHP}. Here we demonstrate, for
the first time, the relevance of this prediction to natural phenomena. The
river level fluctuations of the Danube \cite{JAN} are observed to obey our
prediction, which immediately establishes a generic statistical connection
between turbulence, criticality and company growth statistics.
\end{abstract}

J\'anosi and Gallas have analysed statistics of the daily water level
fluctuations of the Danube collected over the major part of the  $20^{th}$
century~\cite{JAN}. In Figure 1
we show their histogram $P(h)$ of seasonally adjusted river height
fluctuations.
The data are plotted in the form $\sigma P$ against $ \left({h-
\overline{h}\over{\sigma_h}}\right)$, where $\overline{h}$ and $\sigma_h^2$ are
respectively the seasonal mean and varience. This is the form that allows
direct comparison of experiment with limit
functions such as the normal distribution. The data are compared with one
such limit function, the universal probability density function $f(x)$ that
gives exactly the thermodynamic
limit distribution of two model quantities - (a) the critical order
parameter fluctuations of the 2D XY model of magnetism and (b) the steady
state width fluctuations of the 2D Edwards-Wilkinson model of interface growth
\cite{PRE}. In the Figure,
$f(x)$ was obtained by fast Fourier transform of the exactly solved
characteristic function \cite{PRE}. An excellent comparison is obtained
even though Fig.1. contains no
fitting parameters.
As discussed in more detail below, $f(x)$ is asymmetric, with the tail for
fluctuations above the mean going as $x \exp(-x)$, allowing a probability
for a large positive fluctuation that is much larger than that predicted by
a normal distribution.
For example, the largest fluctuation of 5 meters, about $5.5 \sigma_h$,
during the 87 years of data collection, is a once per century
event for $f(x)$ while a
Gaussian probability predicts that such an event only
occurs once every $25$ millenia.
Fluctuations below the mean are of much
smaller
amplitude, with the tail of the distribution going as $\exp(-\exp(x))$.
On the semi-log plot, which emphasises the tails of the distribution, the
experimental data coincide with the theoretical function to an excellent
degree. In natural units (see inset) the two show some deviation near the
origin but
again close agreement in the tails.

\begin{figure}
\onefigure[scale=0.5]{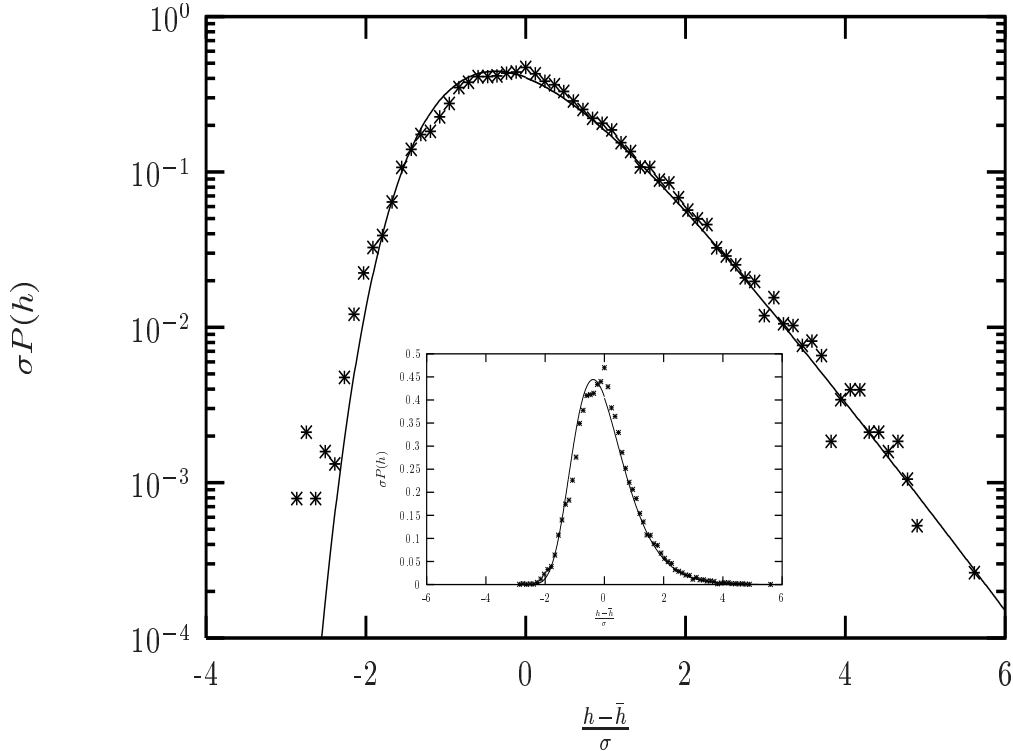}
\caption{
Comparison, with no fitting parameters, between (symbols) the
histogram $P(h)$ of the seasonally adjusted water level
fluctuations of the Danube \cite{JAN} and (line)
the predicted function for complex systems $f(x)$~\cite{PRE}.
The data are plotted as $\sigma P$ against
$\left(\frac{h-\overline{h}}{\sigma_h}\right)$,
where $\overline{h}$ and $\sigma^2$ are respectively
the seasonal mean and variance. For the Danube data $\sigma_h = 83$ cm, so the
fluctuations go out to
nearly 5 metres to the
high water side. {\it
Inset:} same as above, in natural units.
}
\end{figure}

The striking result shown in Fig.1 gives weight to our prediction that
$f(x)$ should occur widely in nature. Specifically, we have predicted that,
if the steady state of a complex system is defined by a global or spatially
averaged quantity $x$, then the latter
will, in many cases, fluctuate according to $f(x)$. This prediction was
motivated by
the observation that $f(x)$ appears to describe fluctuations of an
experimental quantity that is seemingly unrelated to either magnetism or
interface growth: the global power consumption of a turbulent flow in
non-equilibrium steady state \cite{BHP}. Subsequently, $f(x)$ was observed
in a range of numerical models of self-organised criticality, equilibrium
critical behaviour and percolation phenomena \cite{PRL}. With the exception
of turbulence, these examples are all theoretical. However, the Danube data
now provides a radically different experimental system showing
this universal behaviour.
We observe that the close comparison between theory and experiment in Fig. 1 is
obtained without any information of the actual mechanism of water level
fluctuations. In general water level depends on many complex interacting
factors, both
naturally occurring and under human control. These include precipitation,
evaporation, runoff, snowmelt and percolation \cite{DAN}. Such factors must
ultimately determine the shape of the experimental curve in Fig. 1, but the
comparison with $f(x)$ supports the validity of an
alternative phenomenological description, in which the Danube water level is
regarded as a direct and global measure of water transfer over the river basin.

The data collapse shown in Fig.1 also supports a
conjecture made by J\'anosi and Gallas~\cite{JAN} that was motivated by
work of Stanley et al on the statistics of company growth~\cite{STA}.
J\'anosi and Gallas showed that the time
series for the Danube data and that for company growth data share common
statistical properties. They therefore proposed that these
are representative of a wider class of
complex systems~\cite{JAN}. Figure 1 indeed suggests a generic connection
with turbulence, critical phenomena and interface growth.

Returning now to the details of the data collapse:
a limit distribution for a global quantity $y$ is calculated by considering
the distribution of $x = \left({y-\overline{y}\over{\sigma_y}}\right)$ in
the limit of infinite system size (the thermodynamic limit). Here
$\overline{y}$ is the mean and $\sigma_y$ the standard deviation. In an
uncorrelated or weakly correlated system, the central limit theorem
applies, and the quantity $x$ will obey a normal or Gaussian distribution
(an example is the total energy of an ideal gas in the
canonical ensemble~\cite{WAN}). However, in a
strongly correlated system, the central limit theorem can break down, which
means that the limit function can be non-Gaussian. We define $f(x)$ as the
non-Gaussian thermodynamic limit distribution that has been shown to
exactly describe the two model quantities (a) and (b) listed
above~\cite{PRE}. The probability density function $f(x)$ can be obtained by
numerical Fourier transform of its exactly known characteristic function
$\phi(t)$.
We present here a straightforward derivation of $\phi(t)$ which
is useful to illustrate the relationship of $f(x)$ with other standard
distributions.

Consider a global quantity $S = \sum_{\bf n} s_{\bf n}$, where ${\bf n}$ is
a $d$-dimensional vector of integer elements $\pm 1 \dots \infty$ and each
$s_{\bf n}$ is a gamma variable with probability density function
\begin{equation}
\label{gamma}
g(s_{\bf n}) = \frac{a_{\bf n}^\gamma}{\Gamma(\gamma)} e^{-a_{\bf n} s_{\bf
n}} s_{\bf n}^{\gamma - 1}.
\end{equation}
Here $\gamma = \frac{1}{2}$, $a_{\bf n} = |n|^m$ and $m$ is a positive
integer, specified below.

The logarithm of the characteristic function $\psi_{\bf n}(t)$ of $s_{\bf
n}$ may be written as an expansion in cumulents $\kappa_r(s_{\bf n})$:
\begin{equation}
\log \psi_{\bf n}(t) = \sum_1^{\infty} \frac{(it)^r}{r!} \kappa_r(s_{\bf n}),
\end{equation}
\noindent
where the $r$th cumulent, $\kappa_r(s) =
\gamma (r-1)! a_{\bf n}^{-r}$~\cite{KIM}. Using the property that the gamma
variables are statistically independent, the characteristic function of the
compound variate $S$ is the product of the contributions for each ${\bf
n}$:
$\Psi(t) = \prod_{\bf n} \psi_{\bf n}(t)$, so the
$r$th cumulent of $S$ is simply the sum of the $\kappa_r(s_{\bf n})$ :
\begin{equation}
\kappa_r(S) = \frac{1}{2} (r-1)! \sum_{\bf n} \left(\frac{1}{|n|^m}\right)^{r}.
\end{equation}
The limit function is obtained by normalising the variate $S$ by its
standard deviation $\sigma_S$. The $r$th cumulent of $x = S/\sigma_S$ in
the expansion of the characteristic function $\phi(t)$ is:
\begin{equation}
\label{cumulent}
\kappa_r(x) = \frac{
\frac{1}{2} (r-1)! \sum_{\bf n} (\frac{1}{|n|})^{rm}}
{\left(\frac{1}{2} \sum_{\bf n} (\frac{1}{|n|})^{2m}\right)^{r/2}}.
\end{equation}
\noindent
The second, third and fourth cumulents are respectively the variance,
skewness and defect of kurtosis of $x$, and the full probability density
function of $x$ can be found by inverse Fourier transform of its
characteristic function. For general $m$ and $d$ the shape of the
probability density function
depends on the relative importance of the small $|n|$ and large $|n|$
contributions. For small $d$ and large $m$  only the $|n| = 1$ contribution is
important. There are only $2d$ contributions
for  $|n| = 1$ and the function therefore
tends to a $\chi^2$ distribution for $2d$ degrees of freedom.
For large $d$ and small $m$ the function tends to a normal
distribution as many equivalent large $|n|$ contributions dominate the sum
and the second cumulent becomes large compared to all the others (which is
essentially the scenario in the central limit theorem).

In the magnetic model discussed in the abstract, the global quantity, the
magnetization per spin, is the sum of contributions
from $N$ highly correlated spins, but this can be re-cast as contributions
from $N$ independent spin wave modes. The probability density may be
derived in the low temperature approximation, where each mode gives a
statistically independent contribution to the global quantity of the form
(\ref{gamma}), but with $a = (J/T)|q|^m$ where $J$ is
the coupling, $T$ the temperature, $m = 2$ and $q$ is the wavevector
defined in a two-dimensional reciprocal space.
Normalization with respect to the standard deviation removes all
dependence on coupling and system size from the problem and results, after
taking the thermodynamic limit,
in the function (\ref{cumulent}) for the case $d = m = 2$. A detailed study
of this function gives the asymptotes described above~\cite{PRE}. For $d =
m = 2$, the shape of $f(x)$ reflects the importance of both small $|n|$ and
large $|n|$ contributions, which correspond to long and short
wavelength spin wave modes respectively.

In general, for $d = m$, the
mean value of $S$ has a logarithmic divergence with system size,
rather than a power law, which one might consider as a
general characteristic of a critical system.
It should be noted, however, that in the $2D-XY$ model the derivation
sketched here relies on a low temperature approximation to the order
parameter. The true order parameter has the same limit function
$f(x)$, although it cannot be decomposed as a sum of statistically independent
contributions: in that case the mean and all other moments of $S$ do
diverge with system size as power laws~\cite{PRE}. The above results
are therefore
characteristic of a critical system. Indeed we believe that
the frequent observation of $f(x)$ in nature is related the fact that
the low temperature calculation, which is really the limit of criticality,
captures the behaviour of a fully critical regime.

Putting $d = m = 1$ in \ref{cumulent} gives the well-known
Fisher-Tippet or Gumbel distribution, $G(x)$, one of the three limit
functions that describes extreme
value statistics~\cite{KIM,GUM}. This result was recently derived by 
Antal {\it et al} and shown to correspond to the width fluctuations 
of a time series showing ``1/f noise''
\cite{ANT} and was confirmed on experimental data.
In the past extremal statistics
have often been used to model data from natural processes including
river floods \cite{GUM,DAN}.
The functions $f(x)$ and $G(x)$ are
qualitatively quite similar, although
$G(x)$ has a true exponential tail for fluctuations
above the mean~\cite{KIM} while $f(x)$ has a pseudo exponential tail
$\sim x \exp(-x)$. Despite this difference in the asymptotes,
$f(x)$ is given to a good approximation
over any physically oberservable range of fluctuations
by  $f(x) \sim G(x)^a$, where $a \approx \pi/2$ \cite{PRL,PRE}.
We find that the fit of the Danube data to a generalized Gumbel
function of this form
is fairly insensitive to $a$, but that
$f(x)$ gives a significantly better fit than $G(x)$.
The precise relation
of the universal fluctuation phenomena discussed here to extremal
statistics remains is an  interesting question that has been
addressed in a very recent preprint by Dahlstedt and Jensen~\cite{DAH}.

In conclusion, probability distributions for global quantities are
generally rather hard to determine experimentally, limiting the usefulness
of the prediction that these will often follow the form $f(x)$~\cite{BHP}.
However, in view of the fact that $f(x)$ describes the Danube data very well,
it would appear reasonable to conclude that water level is, in this case,
effectively a global measure of a complex system at steady state. In view
of this result and that of \cite{JAN}, it would be interesting to test data
for other rivers as well as appropriate financial data for further
experimental evidence of the proposed far-reaching universality.


{\it Acknowledgements:} We thank I. M J\'anosi for generously providing us
with the Danube data. It is also a pleasure to thank L. Berthier, L.
Bocquet, H. Jensen, J.-F. Pinton and Z. R\'acz for stimulating discussions.

\end{document}